\documentclass[12pt]{iopart}
\usepackage{amssymb}

\usepackage{epsfig}
\usepackage{rotating}
\usepackage{portland}
\usepackage{lscape}
\usepackage{sw20lart}
\usepackage[dvips]{tcigraph}

\input{tcilatex}

\begin{document}

\title {Shapiro delay of asteroids on LISA}
\author{Bertrand Chauvineau, Sophie Pireaux, Tania Regimbau\thanks{}}

\begin{abstract}
In this paper, we examine the Shapiro delay caused by the close approach of
an asteroid to the LISA constellation. We find that the probability that
such an event occurs at a detectable level during the time interval of the
mission is smaller than 1 \% 
\end{abstract}


\rotdriver{dvips}

\address{UMR 6162, ARTEMIS\newline
Observatoire de la C\^{o}te d'Azur,\newline
avenue de Copernic,\newline
06130 GRASSE,\newline
FRANCE\newline
Tel: ++33(0)4 93 40 53 62\newline
Fax: ++33(0)4 93 40 53 33}

\ead{bertrand.chauvineau@obs-azur.fr}

\submitto{\CQG}



\section{Introduction}


LISA (Laser Interferometer Space Antenna) [1], a space experiment devoted to
the detection of gravitational waves, is a nearly equilateral triangular
constellation of three spacecraft, the center of mass of which follows an
Earth-like orbit.\ This constellation is located about 20 degrees behind
Earth, and the distances between spacecraft are planed to be of the order of
5 millions kilometers. The spacecraft exchange optical laser beams, and the
oscillations of the distances between spacecraft are interferometrically
monitored. Thanks to this TDI (Time Delay Interferometry) method,\
gravitational waves signatures are tracked in the $\left[ \sim 10^{-4},\sim
10^{-1}\right] $ $Hz$ frequency domain. To use this method, spacecraft
inter-distances (and their fluctuations) must be known with a precision such
that the gravitational field of the Sun has to be modeled in a relativistic
framework [2].\ The analysis of gravitational waves signal provides another
useful way to explore the universe, giving astrophysical and cosmological
information, inaccessible from the electromagnetic window and complementary
to it.

The gravitational field of the solar system is generally modeled including
the Sun and planets only. However, it is known that the terrestrial orbit is
frequently crossed by asteroids [3,4], refered to as geo-cruisers (GC) in
the following. Since LISA is on an Earth-like orbit, close encounters with
GCs are expected to occur, and the gravitational field of a GC passing close
to the LISA constellation can generate a signal in the data.\ 

In a recent paper, Vinet [5] examined the direct action of a GC on the LISA
constellation.\ The author addressed the shift in position of a station due
to the direct action of the asteroid's gravitational field.\ He finds that
this effect leads to a measurable signal if the involved GC passes
sufficiently close to the spacecraft concerned.

In this paper, we are interested in another aspect of the interaction
between asteroids and LISA.\ If a GC passes close to the segment joining two
spacecraft, its gravitational field affects the light-distance between them
by Shapiro delay on the laser beam used for that very measurement. According
to the close encounter parameters, this signal can have a duration such that
it falls in the LISA frequency domain.\ Hence, this signal has to be
distinguished from the expected extra-solar system gravitational wave signal
to be tracked in the data.\ The aim of the present study is to examine if
such an (impulsive) event is likely to occur. We find that, while the effect
is effectively measurable for sufficiently massive GCs passing close to the
light beam, the probability of the occurence of such an event at a
detectable level, during the time interval of the mission, is quite
negligible.


\section{Conditions for an asteroid encounter to cause a
relevant Shapiro delay}


Let us consider an GC passing close to the light beam linking two LISA
spacecraft A and B. The spacetime geometry in which the beam propagates can
be formally written as 
\begin{equation*}
g_{\alpha \beta }=\eta _{\alpha \beta }+h_{\alpha \beta }^{\text{(S.S.)}%
}+h_{\alpha \beta }^{\text{(ast.)}}+h_{\alpha \beta }^{\text{(G.W.)}}
\end{equation*}
where $\eta _{\alpha \beta }=diag\left( -1,+1,+1,+1\right) $ is the
Minkowsky metric, $h_{\alpha \beta }^{\text{(S.S.)}}$ the part of the
gravitational field due to the Sun and planets, $h_{\alpha \beta }^{\text{%
(ast.)}}$ the part due to the close asteroid, and $h_{\alpha \beta }^{\text{%
(G.W.)}}$ the part of the gravitational wave. The gravitational wave term
induces a change in the distance between the two stations of the order of 
\begin{equation*}
\delta L^{\text{(G.W.)}}\sim hL
\end{equation*}
where $h$ is the characteristic amplitude of $h_{\alpha \beta }^{\text{(G.W.)%
}}$, and $L$ the distance between A and B. On the other hand, $h_{\alpha
\beta }^{\text{(ast.)}}$ is of the order of $2Gm/(rc^{2})$, where $m$ is the
asteroid mass.\ The close approach induces a Shapiro time delay $\delta t$
in the flight time of the photon, hence a change in the light-distance given
in [6] 
\begin{equation*}
\delta l\approx c.\delta t\approx \frac{4Gm}{c^{2}}\ln \left( \frac{%
4r_{A}r_{B}}{\Delta ^{2}}\right) ,
\end{equation*}
where $r_{A}$ (resp.$\;r_{B}$) is the distance between the GC and spacecraft
A (resp.\ B) and $\Delta $ the distance between the GC and the segment
joining the spacecraft A\ and B. Let $b$ be the impact parameter of the
encounter (minimum value of $\Delta $ during the approach).\ In the case
where $b<<L$ (we will find that it is a necessary condition for the signal
to be observable), it is easy to see that $\delta L^{\text{(ast)}}$, the
maximum possible value for $\delta l$, satisfies 
\begin{equation*}
\delta L^{\text{(ast.)}}\lesssim  \frac{8Gm}{c^{2}}\ln \frac{L}{b}.
\end{equation*}


\subsection{Conditions on the amplitude}


Let $H_{\min }$ be the smallest value of $h$ accessible to the experiment.
The necessary condition for the GC to generate a gravitational signal with
sufficient amplitude to be detectable (see the next sub-section for the
necessary condition related to the frequency domain) writes \ 
\begin{equation*}
\frac{8Gm}{c^{2}}\ln \frac{L}{b}\gtrsim H_{\min }L.
\end{equation*}
Let $\rho $ and $D$ be the density and the (mean) diameter of the considered
asteroid respectively.\ The above unequality leads to the following
condition on the impact parameter 
\begin{equation}
b\lesssim L\exp \left\{ -\frac{3}{4\pi }\frac{c^{2}}{G\rho D^{3}}H_{\min
}L\right\}   \label{detect}
\end{equation}
for the asteroid signal to be observable. This gives, numerically, 
\begin{equation}
b\lesssim \left( 5.10^{6}\text{ km}\right) \exp \left\{ -8000.\frac{H_{\min }%
}{10^{-20}}\left( \frac{\rho }{2\text{ g/cm}^{3}}\right) ^{-1}\left( \frac{D%
}{1\text{ km}}\right) ^{-3}\right\}   \label{cond1}
\end{equation}
where we have taken $L=5.10^{6}$ km, the average inter-distance between
spacecraft. Let us consider an GC of $10$ $km$ in diameter (resp. $15$ $km$%
).\ One finds (with $\rho =2$ $g/cm^{3}$ and $H_{\min }=10^{-20}$) $%
b\lesssim 1700$ $km$ (resp. $470000$ $km$). We note that for a $8$ $km$
diameter asteroid, the impact parameter should be smaller than $1$ $km$,
that is smaller than the asteroid radius, which means that the beam would be
occulted.


\subsection{Conditions for the signal to fall in LISA's frequency
interval}

The characteristic time of the encounter is given by $\tau \sim b/V$, where $%
V$ is the relative velocity of the GC with respect to LISA's center of mass.
This means that the fundamental frequency, in the Fourier representation of
the signal, is of the order of $V/b$. Then, a second necessary condition is
that $V/b$\ should be inside the frequency interval of LISA for the signal
to be detectable.\ Simulations with fictitious impulsive signals (i.e.
signals with limited duration), the duration of which range from $1$ to $%
10^{5}$ $s$, confirm that $H_{\min }$ is always $\geq 10^{-20}$, and that
the detection is not efficient outside the interval $\left[ 10\text{ }%
s,10^{4}\text{ }s\right] $\ (see appendix I).

Since one should have $\tau $ in the time interval $\left[ 10\text{ }s,10^{4}%
\text{ }s\right] $, $b$ has to satisfy the additional condition 
\begin{equation}
(10\text{ }s).V\lesssim b\lesssim (10^{4}\text{ }s).V  \label{cond2}
\end{equation}
besides condition (\ref{cond1}). Since $V\sim 15$ $km/s$, (\ref{cond2})
leads to 
\begin{equation}
150\;km\lesssim b\lesssim 150\;000\;km,  \label{cond3}
\end{equation}
one sees that only a small number of GCs will effectively be relevant to
LISA at $H_{\min }=10^{-20}$, verifying both conditions (\ref{cond1}) and (%
\ref{cond2}). Indeed, only asteroids larger than $9$ $km$ in diameter can
generate a signal at a detectable level with characteristic encounter times
in this interval.\ From astronomical observations, only about $15$ GCs are
larger than $9$ $km$ in diameter [4]. If one takes $\rho =2.7$ $g/cm^{3}$,
this limit in diameter becomes $8$ $km$, and about $20$ GCs are larger than $%
8$ $km$ in diameter.


\section{Probability of a relevant encounter}


Let $n\left( \geq D_{0}\right) $ be the mean number density of GCs with a
diameter $D\geq D_{0}$ in the neighbourhood of the Earth orbit. Let $V$ be
the mean relative velocity of GCs and the Earth.\ The number of GCs, of
diameter larger than $D_{0}$, passing at a distance between $b$ and $b+db$
from the segment [A,B] (with $b<<L$) during a time interval $dt$, is of the
order of $2n\left( \geq D_{0}\right) .L.db.V.dt$. Let $T_{\text{LISA}}$ be
the duration of the LISA mission. From eq. (\ref{detect}), the condition of
detectability by LISA is $D\geq D_{0}$, with 
\begin{equation*}
D_{0}^{3}=\frac{3}{4\pi }\frac{c^{2}}{G\rho }\frac{HL}{\ln \left( L/b\right) 
}.
\end{equation*}
The number $E$ of events observed during the duration of the mission is then
of the order of 
\begin{equation*}
E\sim 6LVT_{\text{LISA}}\int_{b_{\min }}^{b_{\max }}n\left( D\geq
D_{0}\right) db
\end{equation*}
since there are three arms in the LISA configuration. The lower and upper
bounds $b_{\min }$ and $b_{\max }$ are the minimal and maximal values of $b$%
, related to the LISA\ frequency sensitivity curve for impulsive events (\ref
{cond3}).\ The total number of GCs of diameter $\geq D_{0}$ is estimated to
be [4] 
\begin{equation}
N\left( \geq D_{0}\right) \sim 1090\left( \frac{D_{0}}{1\text{ }km}\right)
^{-1.95}.  \label{totnmb}
\end{equation}
To evaluate $E$, only the density number in the vicinity of Earth is needed,
not $N$, the total number of GCs.\ From (\ref{totnmb}) and the estimate
obtained in the appendix II, the mean number of GCs per unit volume (per $%
(AU)^{3}$) in the vicinity of the Earth orbit is 
\begin{equation*}
n\left( \geq D_{0}\right) \sim 94\left( \frac{D_{0}}{1\text{ }km}\right)
^{-1.95}.
\end{equation*}
Then 
\begin{equation*}
E\sim \frac{0.51}{\left( 1\text{ }A.U.\right) }\int_{b_{\min }}^{b_{\max }}%
\left[ \frac{\rho }{2\text{ }g/cm^{3}}\left( \frac{H_{\min }}{10^{-20}}%
\right) ^{-1}\ln \frac{L}{b}\right] ^{0.65}db.
\end{equation*}
We have taken $V=15$ $km/s$ and $T_{\text{LISA}}=3$ $yrs$. The values of $%
b_{\min }$ and $b_{\max }$ were given in (\ref{cond3}), and\ $\ln \left(
L/b\right) $ in the integral varies in the interval $\left[ \sim 3.5;\sim
10.4\right] $.\ The minimal amplitude $H_{\min }$ depends on the
characteristic time of the encounter, hence on $b$, but, as stated before,
it can be bounded by $10^{-20}$. Since $\rho $ is always of the order of $2$ 
$g/cm^{3}$ (for asteroids, it belongs to the interval $\left[ 1.3\text{ }%
g/cm^{3};2.7\text{ }g/cm^{3}\right] $), the number of relevant events during
the LISA mission is bounded by 
\begin{equation*}
E\lesssim 1.65\text{ }10^{-3}
\end{equation*}
which means that the probability to observe one event caused by a Shapiro
delay related to a close GC approach is quite negligible. This number
(probability) becomes $2$ $10^{-3}$ if one takes $\rho =2.7$ $g/cm^{3}$.


\section{Discussion}


The present study leads to the conclusion that GCs will not perturb
significantly the LISA mission through related Shapiro effect.\ While it
could appear that this (not very exciting) result is not a surprise, a
dedicated study was required.\ Indeed, we have shown that the close approach
of one of the largest geo-cruisers is susceptible to result in a detectable
signal.\ The low number of relevant encounters comes from the statistical
aspect of the problem, but not directly from the physical properties
(masses) of asteroids neither from the geometry of the possible encounters.

It is worth pointing out that, since only objects with a diameter larger
than $8\;km$ are relevant, all the corresponding geo-cruisers are known.\
Hence, if a relevant close encounter with LISA occurs, the corresponding
geo-cruiser motion will have been accurately monitored, in such a way that
it should be easy to veto the resulting signal. Consequently, it would be
useless to make extensive templates of asteroid's signals, in order to track
such events during the whole LISA mission.

In reference [5], the direct effect of an asteroid on the motion of a LISA
spacecraft has been addressed.\ A statistical analysis of this direct effect
would be of interest and remains to be made.\ To achieve this study, an
analytical expression relating the impact parameter of the encounter and the
minimal value of the asteroid's diameter for which the direct effect is
detectable is required. Such an expression is not explicitly provided in
[5], but it appears from the curves on fig. 3 of [5] that asteroids with a
size of (say) $100\;m$ lead to a detectable effect for impact parameters in
the interval (\ref{cond3}).\ This shows that the direct effect examined in
[5] is considerably more important compared to the Shapiro effect examined
in the current paper, as the latter requires larger GCs (at least $8\;km$ in
diameter).


\section{Conclusion}


The present study shows that the possibility of detecting an asteroid
through Shapiro delay by the LISA mission :

- concerns only a small number of geo-cruisers (about twenty at best).\ An
occultation of the laser beam occurs before the detection condition is
satisfied for geo-cruisers with a diameter less than $\sim 8$ $km$ ;

- has a very low probability to occur during the time interval of the
mission, at best of the order of some $10^{-3}$.

\pagebreak

\noindent \textbf{Appendix I : Minimal detectable amplitude}

\bigskip

The signal to noise ratio averaged over all sky directions and polarizations
can be expressed as [7] 
\begin{equation*}
\left( \frac{S}{N}\right) ^{2}=2\int_{0}^{\infty }d\nu \frac{S_{h}\left( \nu
\right) }{S_{eff}\left( \nu \right) }
\end{equation*}
where $S_{eff}\left( \nu \right) $ is the effective sensitivity of LISA.\ In
our calculations, we adopt the position noise budget for a standard
Michelson configuration, including the contribution of the galactic binary
WD-WD confusion noise [8].

Let $h(t)=Hf(t)$ be a gravitational signal of duration $T$ and of amplitude $%
H$ (the function $f(t)$ being of amplitude unity). The corresponding
spectral density $S_{h}\left( \nu \right) $ can be expressed as 
\begin{equation*}
S_{h}\left( \nu \right) =H^{2}\left| \stackrel{\sim }{f}\left( \nu ,T\right)
\right| ^{2},
\end{equation*}
$\stackrel{\sim }{f}$ being the Fourier transform of $f$.

Combining the above equations, one obtains for the minimal detectable
amplitude 
\begin{equation*}
H_{\min }=\frac{\left( S/N\right) _{\min }}{2\sqrt{I}}
\end{equation*}
where 
\begin{equation*}
I=\int_{0}^{\infty }d\nu \frac{\left| \stackrel{\sim }{f}\left( \nu ,T\right)
\right| ^{2}}{S_{eff}\left( \nu \right) }.
\end{equation*}
Following the convention adopted in the LISA community, we assumed a
detectability threshold of $\left( S/N\right) _{\min }=5$.

\begin{figure}[b]
\begin{center}
\includegraphics[width=1.0\textwidth]{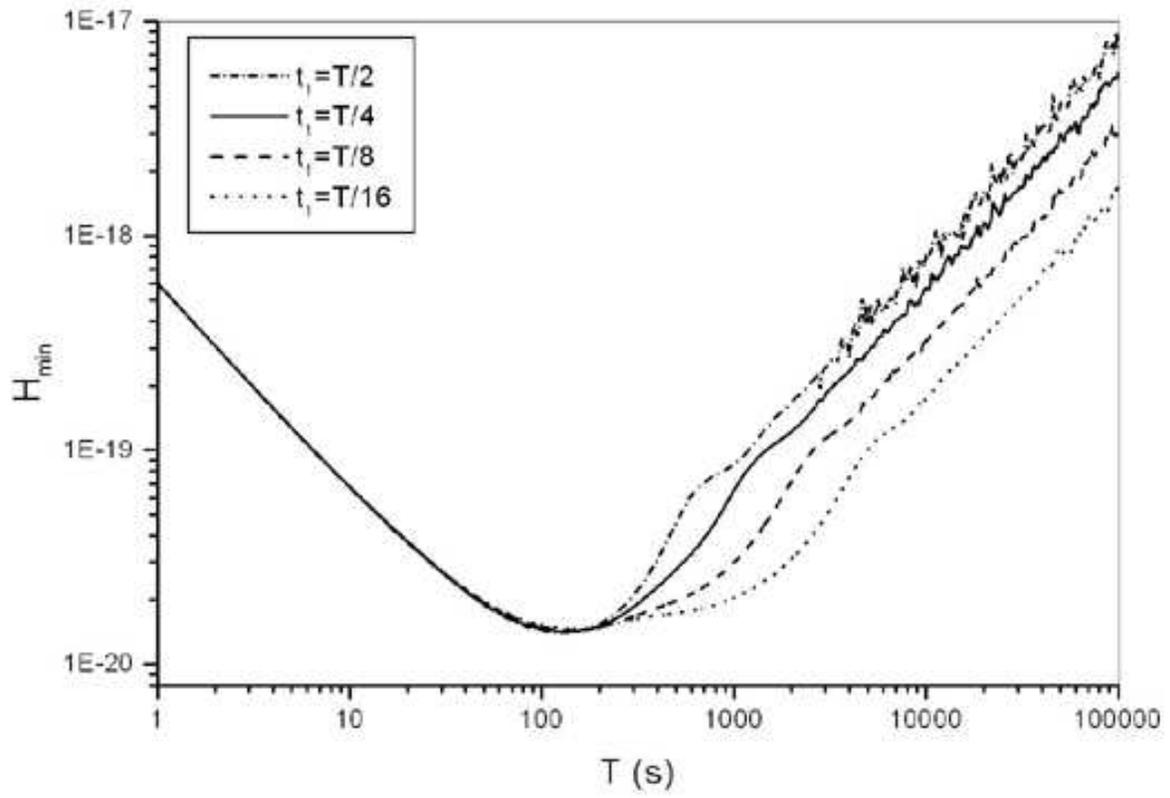}
\end{center}
\caption{minimum signal's amplitude detectable as a function of its
duration, for various fictitious signals with profile described in appendix
I.}
\label{sensitivity_curve}
\end{figure}

For instance, let us consider a (fictitious) signal such that $f(t)$ is zero
outside the interval $\left[ 0,T\right] $, which is unity inside the
interval $\left[ t_{1},t_{2}\right] $, and which is linear in the intervals $%
\left[ 0,t_{1}\right] $ and $\left[ t_{2},T\right] $, in such a way that the
whole signal is continuous. For this signal, Figure \ref{sensitivity_curve} 
exhibits the minimum detectable value $H_{\min }$ as a function of the global duration $T$, for $%
t_{2}=T/2$ and $t_{1}=T/2,\;T/4,\;T/8$ and $T/16$ respectively.\ It confirms
that $H_{\min }$ is always $\geq 10^{-20}$,\ and that this conclusion does
not dependent drastically on the precise signal's profile. It also shows
that the detection is not efficient for a duration outside the interval $%
\left[ 10\text{ }s,10^{4}\text{ }s\right] $, i.e. for durations outside the
LISA frequency interval. More precisely, the detection is not efficient for
a duration $<10\,s$. It can be efficient for a duration $>10^{4}\,s$, but
only for a highly non-symmetric signal, for which the time derivative takes
values significantly larger than $H/T$. However, when a GC passes close to
LISA, the relative velocity of the encounter is nearly constant, so that the
GC gravitational field varies in a very regular and quasi-symmetric way. The
corresponding signal is such that its time derivative is never significantly
larger than $h_{\max }/\tau $ ($h_{\max }$ being the maximum value of the
signal and $\tau $\ its characteristic duration).\ Hence the reasonable
assumption that the detection is not efficient when $\tau $\ is outside the
interval $\left[ 10\text{ }s,10^{4}\text{ }s\right] $.

\bigskip

\noindent \textbf{Appendix II : From total asteroid distribution to volumic
distribution near Earth orbit}

\bigskip

The distribution of GCs with respect to orbital elements ($a,e,i$) is given
in reference [3]. Since one is interested in an order of magnitude estimate
rather than in precise results, let us make the following assumptions and
simplifications :

- the diameter distribution of GCs is independent of the orbital-element
distributions ;

- the distribution in inclination is limited to the interval $\left[
0,i_{\max }\right] $, in which it is uniform.

Let us consider an asteroid with orbital elements $a$\ and $e$. The
probability that this asteroid is at a distance from the Sun in the interval 
$\left[ r,r+dr\right] $, at an arbitrary time, is given by $dP\left(
r,r+dr\right) =2dr/\left( \theta \left| \stackrel{.}{r}\right| \right) $,
where $\theta $ is the period. Then, $\stackrel{. }{r}$ is given by the energy
integral, and one finds 
\begin{equation*}
dP\left( r,r+dr\right) =\frac{dr}{\pi a\sqrt{\left( e\frac{a}{r}\right)
^{2}-\left( 1-\frac{a}{r}\right) ^{2}}}.
\end{equation*}
Let $p\left( a,e\right) $ be the density distribution in $a$ and $e$, so
that $p\left( a,e\right) da.de$ is the probability that a GC, arbitrarily
chosen in the population, has its semi-major axis and eccentricity in the
intervals $\left[ a,a+da\right] $ and $\left[ e,e+de\right] $, respectively.
Using the assumption on inclination distribution, one finds that the number
density of GCs on an Earth orbit is related to the total population $N$ by 
\begin{equation*}
n=\frac{N}{4\pi ^{2}\sin i_{\max }}\int_{\left\{ a,e\right\} }\frac{p\left(
a,e\right) da.de}{a\sqrt{e^{2}a^{2}-\left( 1-a\right) ^{2}}}
\end{equation*}
where one has replaced $r$ by unity ($1$ $A.U.$). In this expression, $a$ is
expressed in $A.U.$, $n$ in $\left( A.U.\right) ^{-3}$ and $\left\{
a,e\right\} $ is the integration domain in the $a-e$ plane. From [3], let us
consider that the integration domain is bounded by $0.5<a<3$ and $0.2<e<1$.\
Besides, for an asteroid to be a GC, one has necessarily $a\left( 1-e\right)
<1<a\left( 1+e\right) $. Since one is only interested in an order of
magnitude, let us repace $p\left( a,e\right) $ by its mean value $%
\left\langle p\left( a,e\right) \right\rangle =\left( 1+\ln 1.44\right)
^{-1}\sim 0.7328$ in the integral. One finds 
\begin{equation*}
n=N\frac{\left\langle p\right\rangle K}{4\pi ^{2}\sin i_{\max }}
\end{equation*}
with $K=3\pi /10+\arcsin \left( 2/3\right) +2/3\ln \left[ \left( 3+\sqrt{5}%
\right) /2\right] \sim 2.3138$. From [3], a reasonnable value for $i_{\max }$
is $30$ degrees. This leads to 
\begin{equation*}
n\left( \geq D_{0}\right) \sim 0.086\;N\left( \geq D_{0}\right) .
\end{equation*}
This is the mean density (per $\left( A.U.\right) ^{3}$) of GCs with
diameter $D\geq D_{0}$, at one astronomical unit from the Sun.

\pagebreak

\clearpage

\bigskip
\pagebreak

\clearpage

\noindent \textbf{Acknowledgements}

\bigskip

We are particularly grateful to P.\ Michel and A.\ Morbidelli, from
Observatoire de la C\^{o}te d'Azur, and J.\ S.\ Stuart, from MIT Lincoln
Laboratory, who provided us with useful information on GCs.

\bigskip

[1] LISA : \textit{a cornerstone mission for the observation of
gravitational waves}, System and Technology Study Report (2000).

[2] B. Chauvineau, T.\ Regimbau, J.-Y.\ Vinet, S.\ Pireaux, Phys. Rev. D 
\textbf{72}, 122003 (2005).

[3] S. N.\ Raymond et al, Astron. J. \textbf{127}, 2978 (2004).

[4] J. S.\ Stuart, R.\ P.\ Binzel, Icarus \textbf{170}, 295 (2004).

[5] J.-Y.\ Vinet, Class.\ Quant.\ Grav. \textbf{23}, 4939 (2006).

[6] S. Weinberg, \textit{Gravitation and Cosmology} (J. Wiley and Sons, New
York, 1972).

[7] J.A.\ de Freitas Pacheco, C.\ Filloux, T.\ Regimbau, Phys. Rev. D 
\textbf{74}, 023001 (2006).

[8] S.L.\ Larson, \textit{Online sensitivity curve generator},

located at http://www.srl.caltech.edu/shane/sensitivity/; S.L.\ Larson,
W.A.\ Hiscock, R.W.\ Hellings, Phys. Rev. D \textbf{62}, 062001 (2000).

\end{document}